\documentclass[prl,showpacs,preprintnumbers,twocolumn]{revtex4}

\usepackage{graphicx}
\usepackage{dcolumn}
\usepackage{float}

\begin{document}

\title{\boldmath The electron-boson spectral density function of underdoped Bi$_2$Sr$_2$CaCu$_2$O$_{8+\delta}$ and YBa$_2$Cu$_3$O$_{6.50}$ \unboldmath}

\author{J. Hwang}
\email[Corresponding author:]{jhwang@pusan.ac.kr}
\affiliation{Department of Physics, Pusan National University, Busan 609-735, Republic of Korea}

\date{\today}


%
%
\begin{abstract}
We investigate the electron-boson spectral density function, $I^2\chi(\omega,T)$, of CuO$_2$ plane in underdoped Bi$_2$Sr$_2$CaCu$_2$O$_{8+\delta}$ (Bi-2212) and underdoped YBa$_2$Cu$_3$O$_{6.50}$ (Y-123) using the Eliashberg formalism. We apply a new (in-plane) pseudogap model to extract the electron-boson spectral function. For extracting the spectral function we assume that the spectral density function consists of two components: a sharp mode and the broad Millis-Monien-Pines (MMP) mode. We observe that both the resulting spectral density function and the intensity of the pseudogap show strong temperature dependences: the sharp mode takes most spectral weight of the function and the peak position of the sharp mode shifts to lower frequency and the depth of pseudogap, $1-\tilde{N}(0,T)$, is getting deeper as temperature decreases. We observe also that the total spectral weight of the electron-boson density and the mass enhancement coefficient increase as temperature decreases. We estimate fictitious (maximum) superconducting transition temperatures, $T_c(T)$, from the extracted spectral functions at various temperatures using a generalized McMillan formula. The estimated (maximum) $T_c$ also shows a strong temperature dependence; it is higher than the actual $T_c$ at all measured temperatures and decreases with temperature lowering. Since as lowering temperature the pseudogap is getting stronger and the maximum $T_c$ is getting lower we propose that the pseudogap may suppress the superconductivity in cuprates.

\end{abstract}


\maketitle


\section{Introduction}

Since the discovery of high temperature superconductivity in the copper oxides\cite{bednorz86} the copper oxides have been studied intensively by numerous condensed matter experimentalists and theoreticians. This group of materials shows a rich, unique and interesting temperature-doping phase diagram. Most areas of the phase diagram are not understood clearly yet. Especially, the pseudogap region has been attracted by many researchers because this region might potentially contain the key properties to answer the long standing unsolved ultimate question: the origin of the microscopic glue to form the Cooper pairs in the material. In the phase diagram underdoped material undergoes from the pseudogap state to the superconducting state through only cooling process. So the two regions should be related very closely to each other. However, there have been two completely opposite scenarios proposed; while some believe that the pseudogap supports the superconductivity through preforming electron-electron pairs, others believe that the pseudogap opposes the superconductivity as a competing order parameter.

To investigate the materials in the pseudogap region one should include the pseudogap in his/her model. Pseudogap is a partial reduction in the electronic density of states (DOS) around the Fermi energy\cite{timusk99}. Recent study\cite{hwang08} showed that the pseudogap could be extracted from ab-plane optical spectra of underdoped cuprates and proposed a new reliable model for the pseudogap. According to the new pseudogap model the density of states loss in the pseudogap is recovered just above the pseudogap energy within the energy scale of pseudogap. We also believe that the boson involved in the electron-boson spectral density function, $I^2\chi(\omega)$, is due to spin fluctuations, where $I$ is the electron-magnetic boson coupling constant and $\chi(\omega)$ is the imaginary part of the spin susceptibility, which can be measured directly by inelastic neutron scattering experiment. We use the Eliashberg formalism to obtain the electron-boson interaction from optical spectra, particularly the optical self-energy\cite{hwang04}. Optically extracted electron-boson spectral density function of underdoped Y123-orthoII has been appeared in two references\cite{hwang06,hwang08a}. In those two references the authors used a constrained electron-boson spectral density function, which consisted of a sharp Gaussian mode with a fixed mode frequency and a broad MMP mode. In one of the two references the authors also applied the new pseudogap model to extract the electron-boson spectral function and got better fits and more reliable results\cite{hwang08a}.

Here we use the new pseudogap model\cite{hwang08} to extract the electron-boson spectral density from optical spectra of two underdoped copper oxides (cuprates): Bi-2212 with $T_c$ = 69 K and Y-123 with $T_c$ = 59 K. For this analysis we set the peak position of the sharp mode as a free parameter, i.e., we impose less constraint on the electron-boson spectral function compared with earlier work\cite{hwang08a} (i.e., one more free parameter) and consequently get better fitting quality. The data and fits for Bi-2212 ($T_c$ = 69 K) and Y-123 ($T_c$ = 59 K) are shown in figure \ref{fig1} and \ref{fig3}, respectively. This new approach gives some interesting and new results. We observe that the depth (or strength) of pseudogap increases almost linearly and the peak in the extracted electron-boson spectral function shifts linearly to lower frequency as temperature decreases. We estimate a fictitious (maximum) superconducting transition temperature $T_c$ using the extracted electron-boson spectral function at measured temperatures and a generalized McMillan formula\cite{mcmillan68,williams89,carbotte90}. As a result, interestingly, the estimated (maximum) $T_c$ decreases with temperature lowering. From these temperature dependent trends in the pseudogap and the maximum $T_c$ we conclude that the pseudogap may suppress the superconductivity in this material. The structure of the paper is as follows. In the next section we introduce a general formalism used for extracting the electron-boson spectral density from the optical self-energy. In the further following sections we show and discuss about results obtained from our analysis.

\section{Optical Spectra and Formalism}

Using the new approach we analyzed published ab-plane optical spectra of underdoped Bi-2212 with $T_c$ = 69 K\cite{hwang07} and a-axis optical spectra of underdoped Y-123 with $T_c$ = 59 K\cite{hwang06}. The spectra analyzed were the optical self-energies, which can be defined in an extended Drude model formalism as follows\cite{hwang04}:
\begin{equation}
\sigma(\omega,T) \equiv -i\frac{\omega}{4\pi}[\epsilon(\omega,T)\!-\!\epsilon_H] \equiv i\frac{\omega_p^2}{4\pi} \frac{1}{\omega\!-\!2\Sigma^{op}(\omega,T)}
\label{eq1}
\end{equation}
where $\sigma(\omega,T) \equiv \sigma_1(\omega,T)+i\sigma_2(\omega,T)$ is the complex optical conductivity, $\epsilon(\omega,T)$ is the complex dielectric function, $\omega_p$ is the plasma frequency, which is proportional to the number density of charge carriers, $\epsilon_H$ is the background dielectric constant at high frequency ($\sim$ 2.0 eV)\cite{hwang07}, and $\Sigma^{op}(\omega,T)$ is the complex optical self-energy. For more detailed discussion and information about $\omega_p$ and $\epsilon_H$ one can refer to Ref. \cite{hwang07}. The optical self-energy should be a complex function, $\Sigma^{op}(\omega,T)\equiv \Sigma^{op}_1(\omega,T)+i\Sigma^{op}_2(\omega,T)$, to hold the conditions of causality~\cite{wooten72}. This quantity can carry the information of correlation between charge carriers in the material, like the quasiparticle self-energy. The real part of it is related to the mass renormalization from the correlation, $-2\Sigma^{op}_1(\omega,T) \equiv -\omega_p^2/4\pi\: \mbox{Im}[1/\sigma(\omega,T)] -\omega \equiv [m^*(\omega,T)/m-1]\omega \equiv \omega\lambda(\omega,T)$, where $m^*(\omega,T)$ is the effective mass and $m$ is the bare electron mass and $\lambda(\omega,T)$ is the mass renormalization function. The imaginary part is related to the frequency dependent relaxation time of charge carriers, $-2\Sigma^{op}_2(\omega,T) \equiv \omega_p^2/4\pi\: \mbox{Re}[1/\sigma(\omega,T)] \equiv 1/\tau(\omega,T)$, where $1/\tau(\omega,T)$ is the optical scattering rate. The optical self-energy is different from the quasiparticle self-energy, which can be measured by an angle-resolved photoemission spectroscopy (ARPES)\cite{carbotte05,hwang07b}. The optical self-energy is an averaged quantity over the Fermi surface and contains a two-particle process, which is a more complicated quantity than the single-particle process in the quasiparticle self-energy. The optical scattering rates at normal state for underdoped Bi-2212 and Y-123 are shown in figure \ref{fig1} and \ref{fig3}, respectively.

\begin{figure}
  \centering
 \vspace*{-1.0cm}%
 \centerline{\includegraphics[width=3.8in]{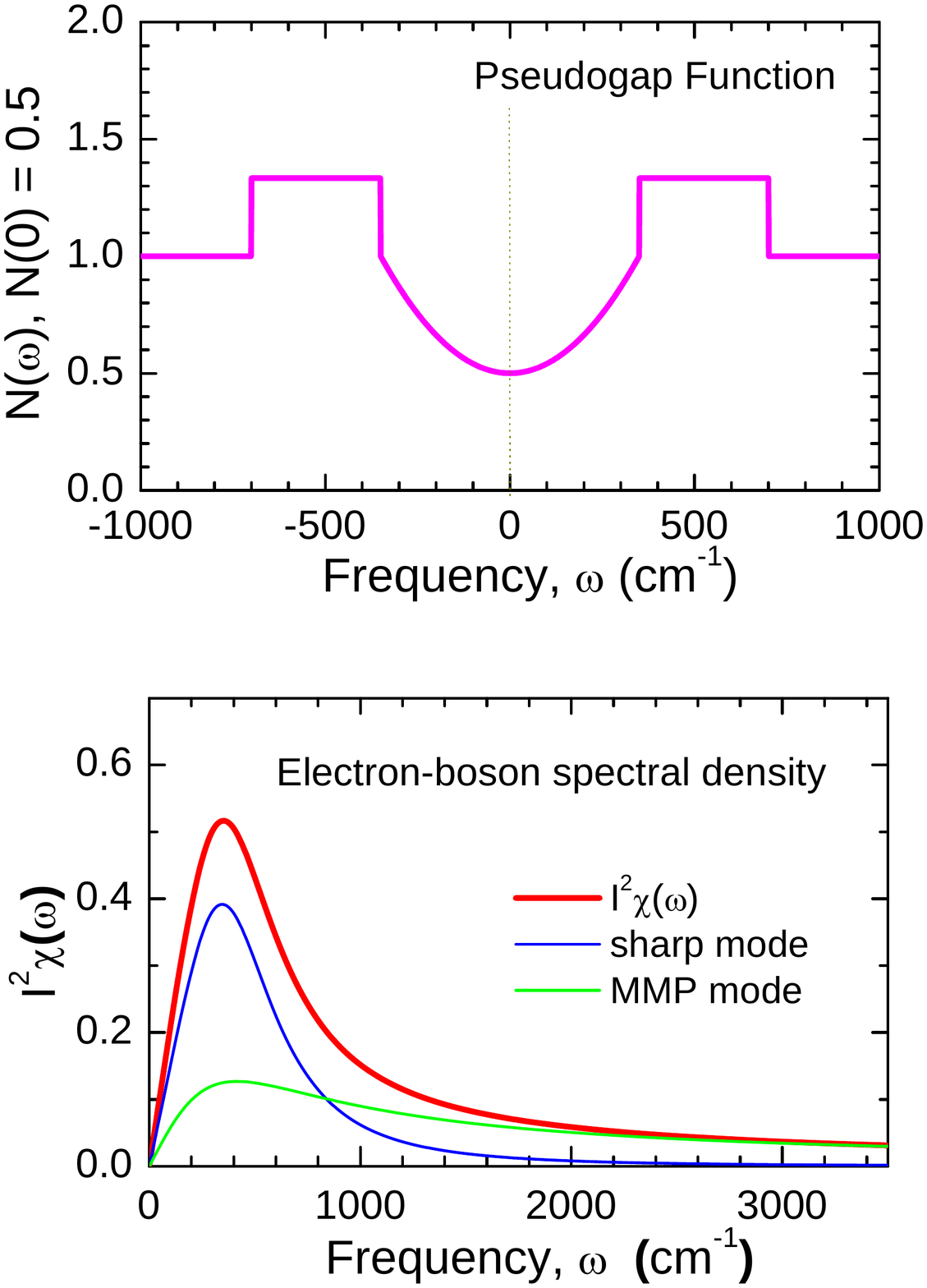}}%
 \vspace*{-1.5cm}%
\caption{(color online) The pseudogap model is displayed in the upper panel as an example (see in the text for a more detailed description). In the lower panel we display the electron-boson spectral density function, which consists of two modes as shown separately in the panel (see in the text for a more detailed description).} \label{fig0}
\end{figure}

To obtain the electron-boson spectral density function, which shows the interactions between charge carriers more explicitly, we use the integral equations, which relate the optical self-energy to the electron-boson spectral density function, derived by Allen\cite{allen71} within an Eliashberg formalism.
\begin{eqnarray}
-2\Sigma^{op}_1(\omega,T)\!\!\!&=&\!\!\!2\!\!\int^{\infty}_{0}\!\!\!\!\!\!d\Omega \:I^2\chi(\Omega,T)\Big{[}\frac{\Omega}{\omega}\ln{\Big{|}\frac{\Omega^2\!-\!\omega^2}{\Omega^2}\Big{|}\!+\!\ln{\Big{|}\frac{\Omega\!-\!\omega}{\Omega\!+\!\omega}\Big{|}}} \Big{]} \nonumber\\
-2\Sigma^{op}_2(\omega,T)\!\!&\equiv&\!\!\frac{1}{\tau(\omega,T)}\!=\!\frac{2\pi}{\omega}\int^{\infty}_{0}\!\!\!\!d\Omega \:I^2\chi(\Omega,T) (\omega\!-\!\Omega)
\label{eq2}
\end{eqnarray}
where $I^2\chi(\Omega,T)$ is the electron-boson spectral density function. Even though the above equations explain important qualitative characteristics of correlated electron systems they are valid for a constant density of states at $T$ = 0 K. So Shulga {\it et al.}\cite{shulga91} extended the formula and obtained a more generalized one, which can be used for material systems at finite temperatures, as follows:
\begin{eqnarray}
\frac{1}{\tau(\omega,T)}\!\!\!&=&\!\!\!\frac{\pi}{\omega}\!\!\int^{\infty}_{0}\!\!\!\!\!\!d\Omega \:I^2\chi(\Omega,T) \Big{[}2\omega \coth{\frac{\Omega}{2T}}
\!-\!(\omega\!+\!\Omega) \coth{\frac{\omega\!+\!\Omega}{2T}} \nonumber \\ &+&(\omega\!-\!\Omega)\coth{\frac{\omega\!-\!\Omega}{2T}}\Big{]}
\label{eq3}
\end{eqnarray}
But this formula is not valid for a non-constant density of states. So we can not apply it for material systems, which have non-constant density of states, like underdoped cuprates. Further generalized formula, which can be applied for material systems with non-constant density of states at finite temperatures, was derived by Sharapov and Carbotte\cite{sharapov05}.
\begin{eqnarray}
\frac{1}{\tau(\omega,T)}\!\!\!&=&\!\!\!\frac{\pi}{\omega}\!\! \int^{+\infty}_{0}\!\!\!\!\!d\Omega I^2\chi(\Omega,T)\int^{+\infty}_{-\infty}\!\!\!\!\!ds[\tilde{N}(s\!-\!\Omega,T)\!+\!\tilde{N}(\Omega\!-\!s,T)]\nonumber \\
&\times&[n_B(\Omega)+f(\Omega-s)][f(s-\omega)-f(s+\omega)]
\label{eq4}
\end{eqnarray}
where $\tilde{N}(\Omega,T)$ is the normalized electronic density of states, with which we can handle non-constant density of states. $n_B(\Omega)=1/(e^{\beta\Omega}-1)$ and $f(\Omega)=1/(e^{\beta\Omega}+1)$ are the Bose-Einstein and Fermi-Dirac occupation numbers, respectively, and $\beta = 1/(k_B T)$, where $k_B$ is the Boltzmann constant and $T$ is the temperature.

For this study we use the most generalized electron-boson formula, Eq. \ref{eq4} since we investigate underdoped cuprates, which have non-constant density of states, i.e., the pseudogap. We also use the new pseudogap model proposed by Hwang {\it et al.}\cite{hwang08}. A more detailed description of the pseudogap model used here is as follows: the shape of pseudogap is shown in the upper panel of figure \ref{fig0}. In the pseudogap model the width of pseudogap ($\Delta_{pg}$) is fixed as 350 cm$^{-1}$ for all temperatures and the electronic density of states loss in the pseudogap near the Fermi energy is recovered completely between $\Delta_{pg}$ and $2\Delta_{pg}$ as shown in the figure\cite{hwang08}. The normalized density of states can be written as follows:
\begin{eqnarray}
\tilde{N}(\omega,T)\!\!&=&\!\!\tilde{N}(0,T)\! +\! [1\!-\!\tilde{N}(0,T)]\Big{(} \frac{\omega}{\Delta_{pg}}\Big{)}^2\:\:\mbox{for} \:|\omega|\leq \Delta_{pg} \nonumber \\
&=&\! 1+\frac{2[1\!-\!\tilde{N}(0,T)]}{3} \:\:\:\:\:\:\:\mbox{for}\: |\omega|\in (\Delta_{pg},2\Delta_{pg}) \nonumber \\
&=& \!1 \:\:\:\:\:\:\:\:\:\:\:\:\:\:\:\:\:\:\:\:\:\:\:\mbox{for} \:\:|\omega|> 2\Delta_{pg}
\end{eqnarray}

One example of the model electron-boson spectral density function is also shown in the lower panel of figure \ref{fig0}. The model electron-boson spectral density function consists of two components: an asymmetric sharp mode and the MMP mode\cite{millis90} as shown separately in the figure. It can be written as follows:
\begin{eqnarray}
I^2\chi(\omega,T) = \frac{A_s(T) \:\omega}{[\omega_s(T)]^4+\omega^4}+\frac{A_m(T)\: \omega}{[\omega_m(T)]^2+\omega^2}
\end{eqnarray}
where $A_s(T)$ and $\omega_s(T)$ are proportional to the amplitude and the peak frequency for the sharp mode, respectively. More accurately, the area under the sharp mode is $(A_s/2\omega_s^2) \tan^{-1}(\omega_c^2/\omega_s^2)$ with a cutoff frequency $\omega_c$ and the sharp peak position is $\omega_{peak,s} = \omega_s/3^{1/4}$. $A_m(T)$ and $\omega_m(T)$ are also proportional to the amplitude and the peak frequency for the MMP mode, respectively. The area under the MMP curve is $(A_m/2) \ln|(\omega_m^2+\omega_c^2)/\omega_m^2|$ with a cutoff frequency $\omega_c$ and the MMP peak position is $\omega_{peak,m} = \omega_m$. Here we use $\omega_c$ = 5000 cm$^{-1}$ for Bi-2212 and $\omega_c$ = 3500 cm$^{-1}$ for Y-123.

\section{Results of numerical analysis and discussions}

\begin{figure}
  \centering
 \vspace*{-1.3cm}%
 \centerline{\includegraphics[width=3.8in]{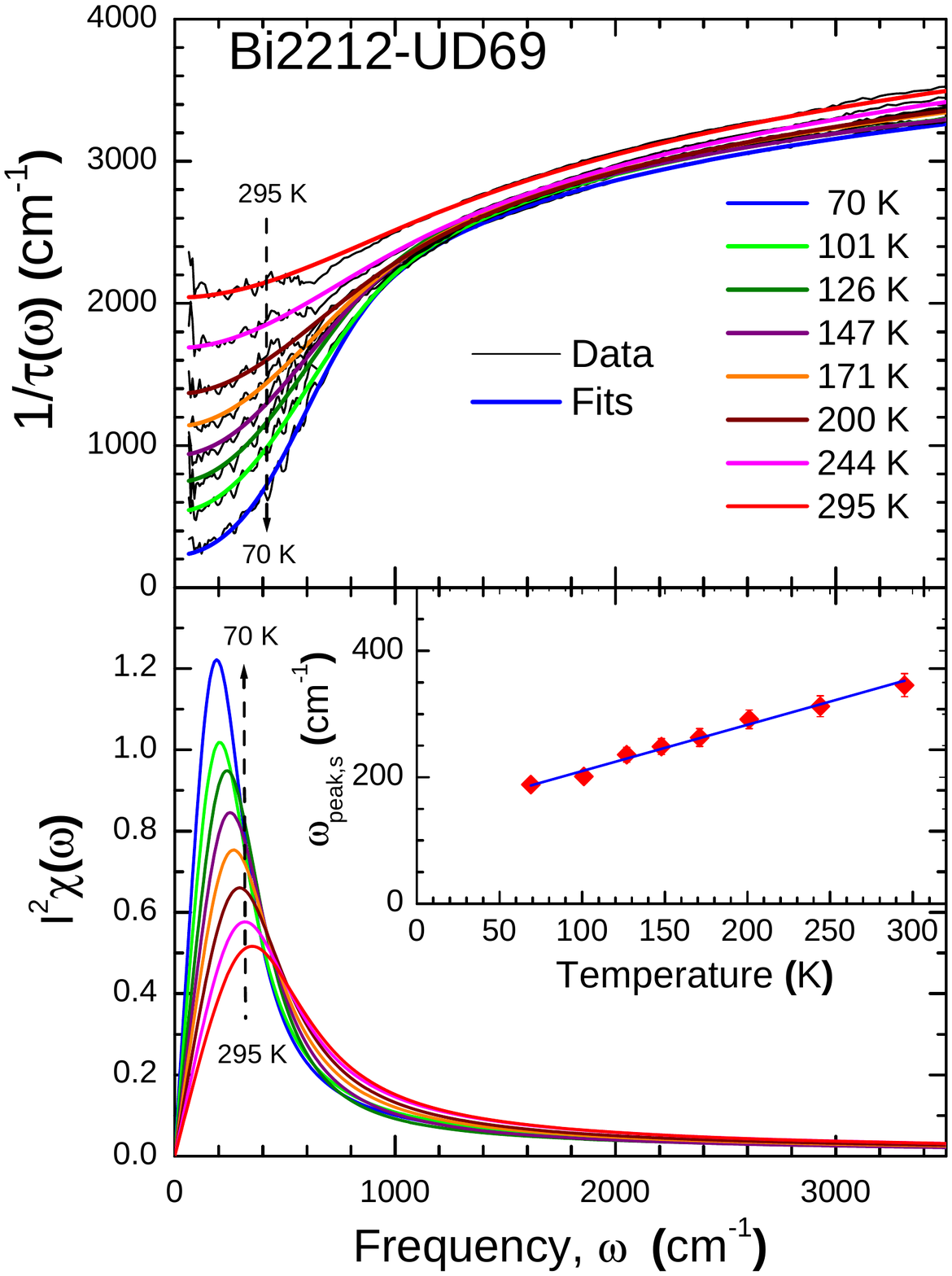}}%
 \vspace*{-1.3cm}%
\caption{(color online) The measured optical scattering rates (thin lines) and corresponding simulated data (thick lines) for underdoped Bi-2212 with $T_c$ = 69 K are shown in the upper panel. The resulting electron-boson spectral functions at various temperatures obtained from the fittings are depicted in the lower panel. In the inset we show the temperature dependent peak position in the electron-boson density function, $\omega_{peak,s}$. The solid line is provided as a guide to the eye.} \label{fig1}
\end{figure}

\begin{figure}
  \centering
 \vspace*{-0.5cm}%
 \centerline{\includegraphics[width=3.8in]{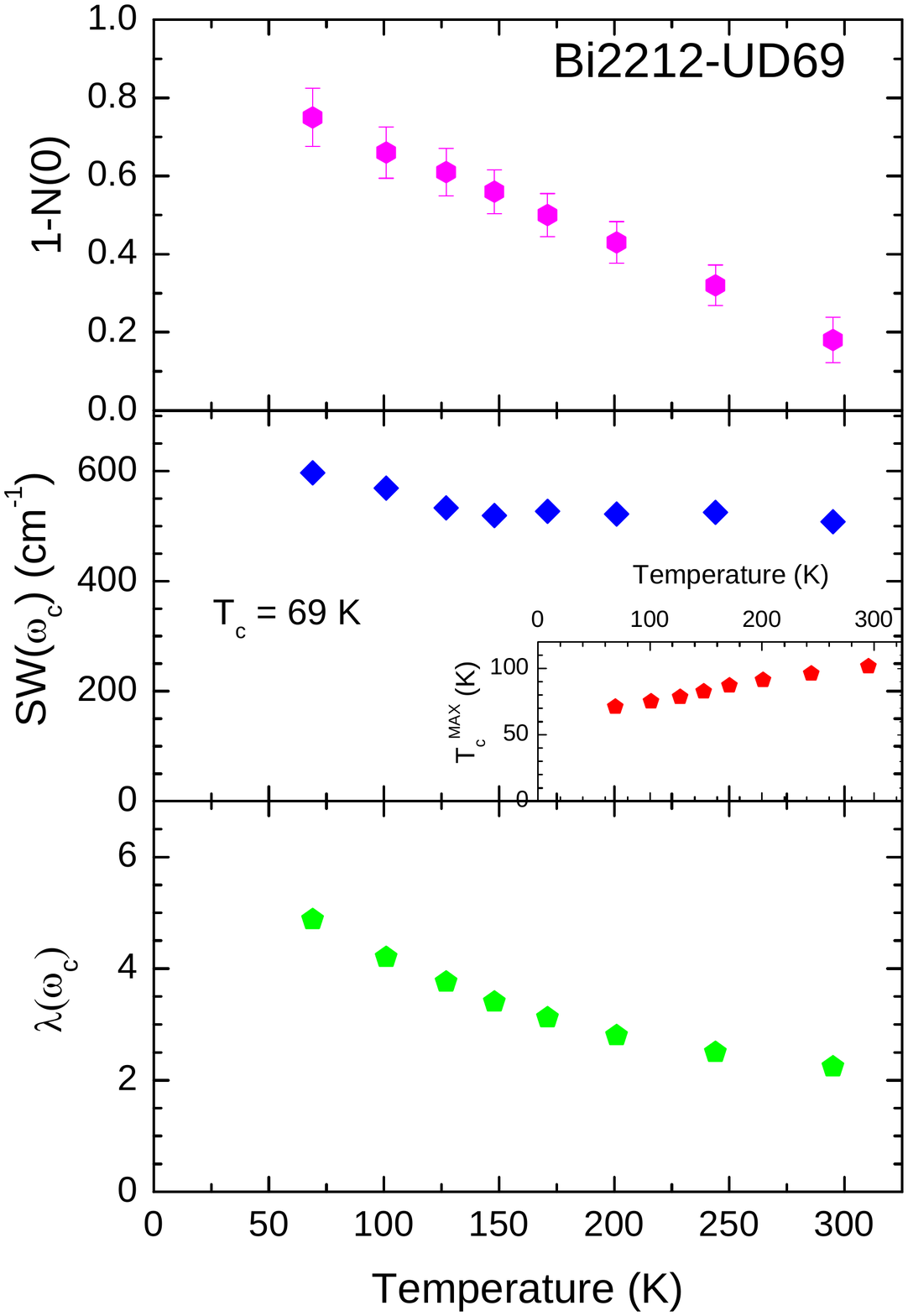}}%
 \vspace*{-1.0cm}%
\caption{(color online) We display the resulting three temperature dependent fitting parameters: the pseudogap depth, 1-$\tilde{N}(0)$, in the top panel, the spectral weight of the electron-boson spectral density in the middle panel, and the mass renormalization factor in the bottom panel. In the inset of the middle panel we also show the estimated temperature dependent $T_c^{MAX}(T)$.} \label{fig2}
\end{figure}

\begin{figure}
  \centering
 \vspace*{-1.0cm}%
 \centerline{\includegraphics[width=3.8in]{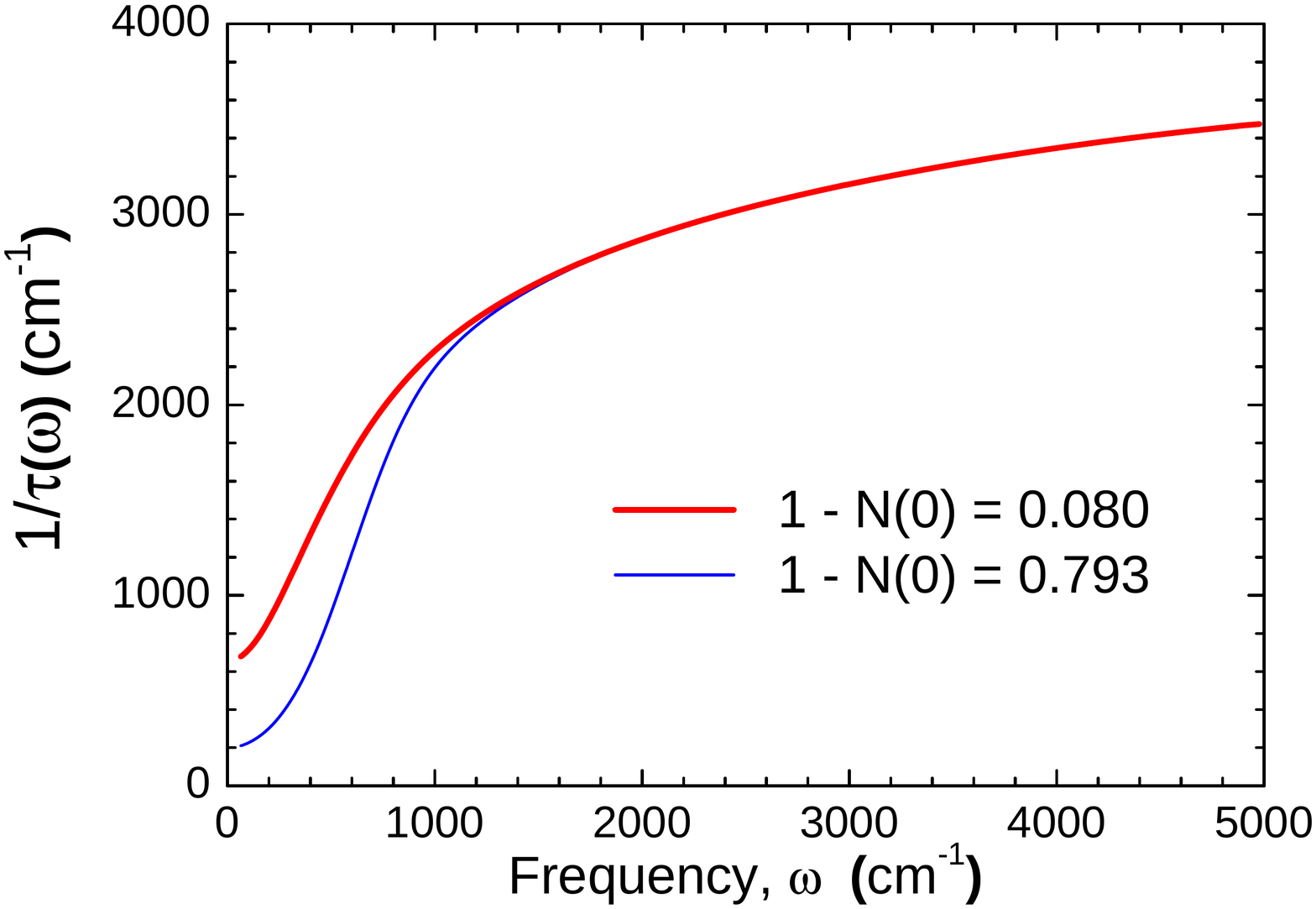}}%
 \vspace*{-0.5cm}%
\caption{(color online) Calculated $1/\tau(\omega)$ using the same $I^2\chi(\omega)$ and two different pseudogap depthes, $1-\tilde{N}(0)$ = 0.080 and 0.792 (see in the text for a more detailed description).} \label{fig2a}
\end{figure}

We use the formalism discussed in the previous section to analyze our two sets of the optical scattering rates, i.e., the imaginary part of the optical self-energy, of underdoped Bi-2212 and Y-123 at various temperatures above $T_c$. In the upper panel of figure \ref{fig1} we display the measured optical scattering rates of underdoped Bi-2212 with $T_c$ = 69 K at various temperatures and corresponding fits obtained by using the formalism and a numerical analysis (i.e., a least square process). For the numerical analysis at a given temperature we fix the temperature ($T$), the shape and width of the pseudogap and the shape of the electron-boson density function (as shown in figure \ref{fig0}) and have five free fitting parameters: the pseudogap intensity (i.e., $1-N(0)$) and the two mode frequencies ($\omega_s$ and $\omega_m$) and the two mode intensities ($A_s$ and $A_m$) of the sharp and MMP modes in $I^2\chi(\omega)$. We assumed the fixed pseudogap width as $\Delta_{pg}$ = 350 cm$^{-1}$ for all temperatures studied, which is a reasonable assumption gotten from an observation of tunneling spectra measured by Renner {\it et al.}\cite{renner98}. As we can see in figure \ref{fig1} the quality of fits is pretty good. We note here that the pseudogap and the sharp peak give different line shapes to the optical scattering rates even though the sharp increase in the scattering rate near 600 cm$^{-1}$ comes from positive contributions of both the pseudogap and the sharp mode \cite{hwang06,hwang08,hwang08a}. So we can obtain the pseudogap strength and the sharp mode frequency separately from the analysis. We will discuss this issue more in detail later in the section. In the lower panel of figure \ref{fig1} we show the resulting electron-boson spectral density function at various temperatures. The temperature dependent peak position of the sharp mode is also shown in the inset of the lower panel. The peak is softened linearly from 370 to 185 cm$^{-1}$ as temperature decreases from 295 to 70 K. This is a similar temperature dependent trend of the sharp mode in the electron-boson density of optimally doped Bi-2212, which has been reported in a reference.\cite{hwang07a} (see figure 2d in the reference).

Results of a further analysis of the extracted $I^2\chi(\omega,T)$ are displayed in figure \ref{fig2}. First of all, most spectral weight of the resulting electron-boson spectral function resides in low frequencies, within 1000 cm$^{-1}$. The spectral weight distribution is qualitatively different from those of optimally and overdoped systems\cite{hwang07a}; for a very overdoped system the spectral weight spreads almost uniformly in a wide spectral range from 0 through 400 meV (see figure 2c in \cite{hwang07a}). In the top panel we show the pseudogap depth at the Fermi surface, $1-\tilde{N}(0)$, as a function of temperature. The pseudogap depth can be a measure of the intensity of pseudogap and increases linearly as temperature decreases, which agrees well with the Fermi arc model of the pseudogap studied by Kanigel {\it et al.}\cite{kanigel06}, equivalently $\tilde{N}(0,T) \propto T/T_c$\cite{hwang08a}. In the middle panel we show the total spectral weight ($SW$) of the electron-boson density spectrum as a function of temperature, which is defined by $SW(\omega_c,T) \equiv \int^{\omega_c}_{0}I^2\chi(\Omega,T)\: d\Omega$, where $\omega_c$ is the cutoff frequency. Here we use $\omega_c$ = 5000 cm$^{-1}$. The spectral weight is almost constant down to 150 K and below the temperature it increases rapidly. In the bottom panel we display the mass renormalization factor, $\lambda(\omega_c,T) = m^*(\omega_c,T)/m -1 \equiv 2\int^{\omega_c}_{0}I^2\chi(\Omega,T)/\Omega \:\:d\Omega$. The mass renormalization factor increases almost linearly from 295 to 150 K and below the temperature it increases more rapidly; there is a slope change near 150 K.

We are able to estimate a fictitious (maximum) superconducting transition temperature, $T_c(T)$, at each measured temperature from the extracted electron-boson spectral density using a generalized McMillan formula\cite{mcmillan68,williams89,hwang08b}. Here we assume that the whole electron-boson spectral density contributes to the fictitious superconducting $T_c$. The generalized McMillan formula can be written as follows \cite{hwang08b}:
\begin{eqnarray}
k_B T_c(T) &\cong& 1.13\:\hbar\:\omega_{\ln}(T)\exp{\Big{[}-\frac{1+\lambda(T)}{g \lambda(T)}\Big{]}} \:\:\:\mbox{or} \nonumber \\
T_c(T) \: &\cong& 1.626 \:\bar{\nu}_{\ln}(T)\:\exp{\Big{[}-\frac{1+\lambda(T)}{g \lambda(T)}\Big{]}}
\label{Tc}
\end{eqnarray}
where $\lambda(T)$ is the mass renormalization factor. $\omega_{\ln}(T)$ is the (logarithmically) averaged boson frequency of the electron-boson density function\cite{allen75}; $\omega_{\ln}(T) \equiv \exp{[(2/\lambda(T))\int^{\omega_c}_{0}\ln{\Omega}\:\:I^2\chi(\Omega,T)/\Omega\:d\Omega]}$. $g$ is an adjustable parameter ($g \in [0,1]$), which may allow us to take the d-wave nature of the superconductivity into account in the formula; the anisotropicity of the d-wave superconducting gap may oppose the superconductivity. When $g$=1, $T_c$ becomes its maximum value, which we will denote as $T_c^{MAX}$; when $g$ = 0, $T_c$ = 0 K. In the lower formula in Eq. \ref{Tc} $\bar{\nu}(\omega_{\ln})$ is in the wavenumber unit, i.e., cm$^{-1}$. So the lower formula might be useful for a practical application in optical studies.

\begin{figure}
  \centering
 \vspace*{-1.3cm}%
 \centerline{\includegraphics[width=3.8in]{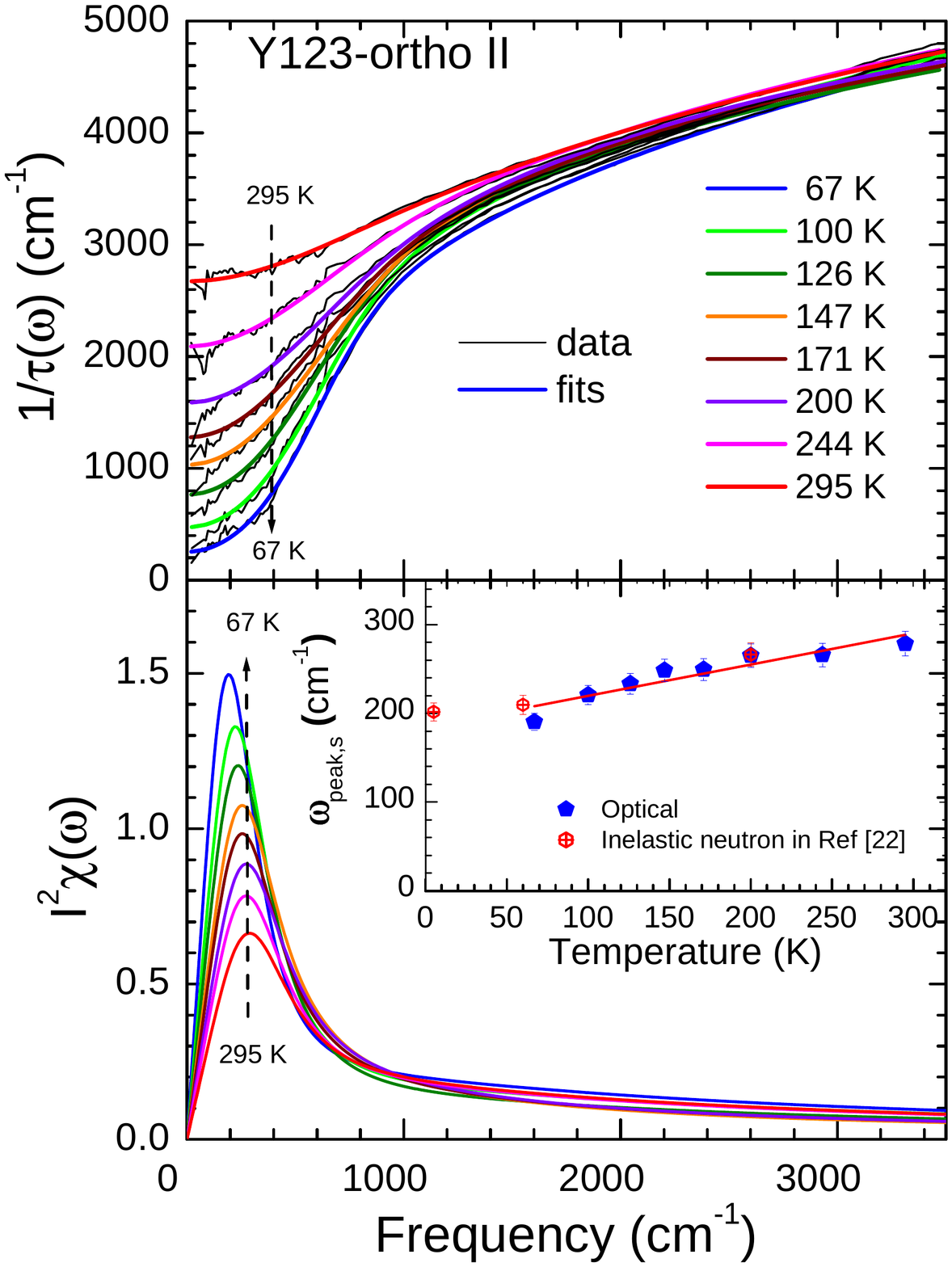}}%
 \vspace*{-1.3cm}%
\caption{(color online) The measured optical scattering rates (thin lines) and corresponding simulated data (thick lines) for underdoped Y-123 with $T_c$ = 59 K are shown in the upper panel. The resulting electron-boson spectral functions at various temperatures obtained from the fits are depicted in the lower panel. In the inset we superimpose the temperature dependent peak position of $I^2\chi(\omega)$, $\omega_{peak,s}$ and a temperature dependent neutron peak extracted from Fong {\it et. al} study. The solid line is provided as a guide to the eye.\cite{fong00,fong00a}} \label{fig3}
\end{figure}

\begin{figure}
  \centering
 \vspace*{-0.5cm}%
 \centerline{\includegraphics[width=3.8in]{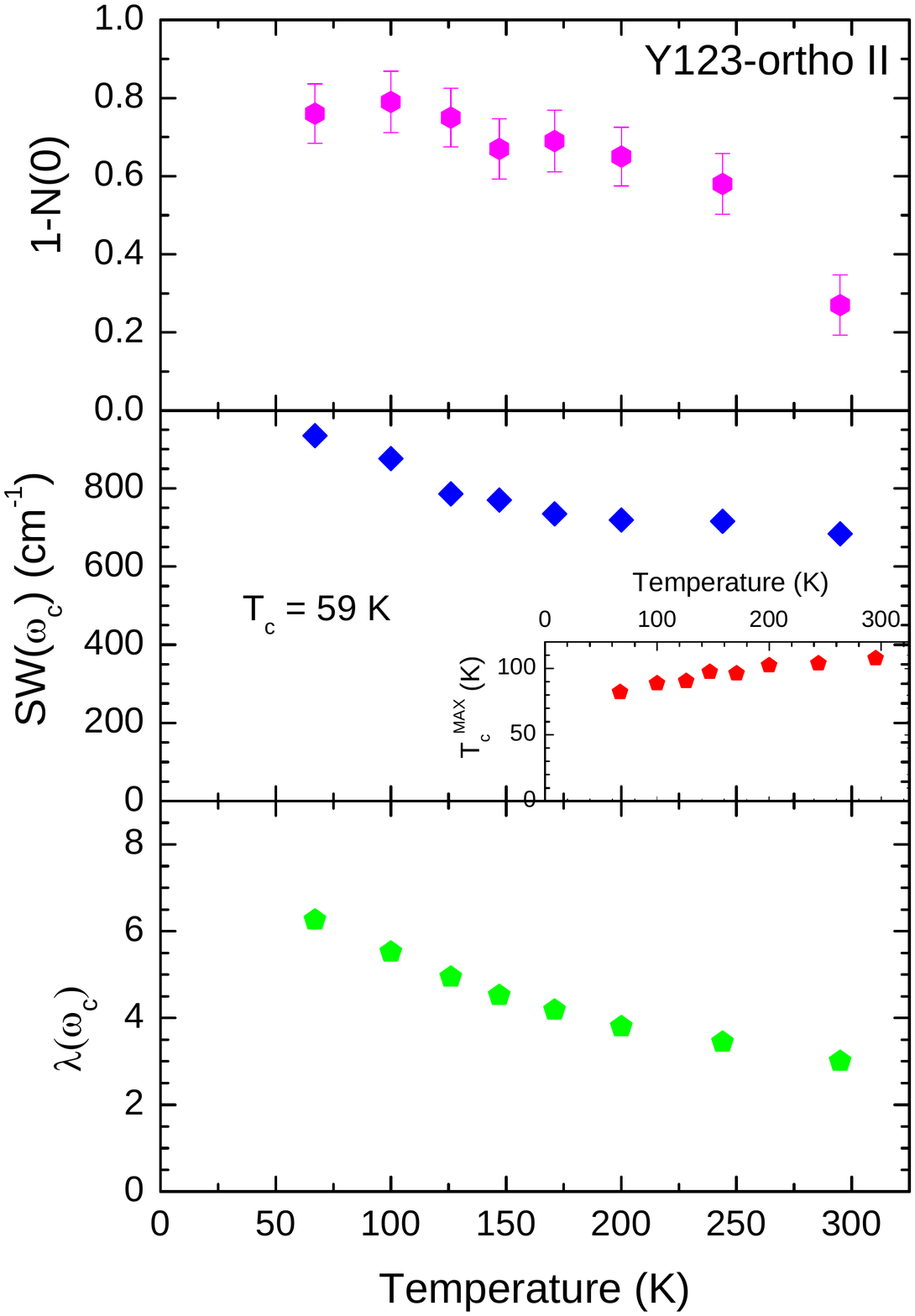}}%
 \vspace*{-1.0cm}%
\caption{(color online) From the fits of the optical scattering rates of underdoped Y-123 sample we obtain and display the resulting pseudogap depth, 1-$\tilde{N}(0)$, in the top panel, the spectral weight of the electron-boson spectral density in the middle panel, and the mass renormalization factor in the bottom panel. In the inset of the middle panel we also display the estimated temperature dependent $T_c^{MAX}(T)$.} \label{fig4}
\end{figure}

We display $T_c^{MAX}(T)$ at various temperatures in the inset of the middle panel of figure \ref{fig2}. Since as temperature decreases the peak position shifts to lower frequency the averaged boson frequency, $\omega_{\ln}(T)$ also deceases. Even though $\lambda(T)$ increases the resulting $T_c^{MAX}(T)$ decreases as lowering temperature as shown in the inset. We will discuss this issue more in detail in the following paragraph. At our lowest temperature $T_c^{MAX}(T = 70 K) \simeq $ 71 K, which is pretty close to the actual $T_c$, 69 K. The peak position of the sharp mode and $T_c^{MAX}$ show a positive correlation; both decrease as temperature is lowered, roughly $\omega_{peak,s} \simeq 4.5 k_B T_c^{MAX}$. $T_c^{MAX}$ is also pretty sensitive to the intensity of the electron-boson density near zero frequency. To demonstrate it we perform the same numerical analysis for underdoped Bi-2212 at $T$ = 70 K with a different electron-boson spectral density model described by the equation below; in this model there is a (spin) gap in the electron-boson density spectrum. The width of the gap is 100 cm$^{-1}$.
\begin{eqnarray}
I^2\chi(\omega) &=& 0 \:\:\:\:\:\:\:\:\:\:\:\:\:\:\:\:\:\:\:\:\:\:\:\:\:\:\:\:\:\:\mbox{for}\:\:\omega < 100 \:\mbox{cm}^{-1} \nonumber \\
&=& \frac{A_s \omega}{\omega_s^4+\omega^4}+\frac{A_m \omega}{\omega_m^2+\omega^2}\:\: \mbox{for}\:\:\omega\geq 100 \:\mbox{cm}^{-1}
\end{eqnarray}
The quality of fit is still pretty good (not shown). We estimate a new $T_c^{MAX}$ by using the generalized McMillan equation and new resulting electron-boson spectral function and get a higher $T_c^{MAX}$, 98 K; it is increased from 71 K to 98 K by 38 \%.

The stronger pseudogap means that the pseudogap gives effectively the more reduction in the optical scattering rate near zero frequency (i.e., the Fermi surface) by reducing the more electronic density of states near the Fermi surface. To see the pseudogap effect on the scattering rate more clearly we calculate two $1/\tau(\omega)$ using the same extracted $I^2\chi(\omega)$ of Bi-2212 at $T$ = 70 K with two different pseudogap depthes, $1-\tilde{N}(0)$ = 0.080 and 0.792. The figure \ref{fig2a} shows results of the calculations. We can see that the deeper (i.e., stronger) pseudogap gives effectively the higher onset scattering edge even though we have the same peak position in $I^2\chi(\omega)$ and the same width of the pseudogap. In fact, both the pseudogap and the sharp peak position contribute positively to the onset frequency in the optical scattering rate; both make the onset frequency larger\cite{hwang06,hwang08,hwang08a}. So for a given onset scattering edge if we have the stronger pseudogap we would have the lower peak position.

We observe that there is a negative correlation between the strength of the pseudogap and $T_c^{MAX}$(T), i.e., as temperature decreases while the pseudogap is getting stronger, the maximum superconducting temperature $T_c^{MAX}(T)$ is getting lower. As we have discussed previously the stronger pseudogap causes effectively the lower peak position of $I^2\chi(\omega)$. So as lowering temperature since the pseudogap strength is getting stronger the peak position shifts to lower frequency. Consequently the averaged boson frequency, $\omega_{\ln}(T)$ deceases. It is clear that the pseudogap gives a negative effect on $T_c^{MAX}(T)$ by shifting the sharp mode to lower frequency through the generalized McMillan formalism. On the other hand the coupling strength, $\lambda(T)$, increases as temperature decreases (see the bottom panel of figure \ref{fig2}). Surprisingly, in spite of the increase of the coupling strength, $\lambda(T)$, the resulting $T_c^{MAX}(T)$ decreases as lowering temperature. This result seems somewhat counterintuitive. However, we may be able to understand the behavior as follows: two contributions from the pseudogap and the coupling compete against each other and the temperature dependent effect of the pseudogap on $T_c^{MAX}(T)$ seems to be larger than that of the coupling strength. In other words, for a given amount of temperature reduction the negative contribution of the pseudogap to $T_c^{MAX}$(T) is larger than the positive contribution of the coupling strength; the pseudogap possibly suppresses the superconductivity in the material. We should note that we consider only the fictitious maximum $T_c$ (i.e., $g$ =1) to get the conclusion. For a real material the parameter $g$ is unknown. As $g$ increases the coupling strength effect on $T_c^{MAX}(T)$ is getting stronger. When $g \simeq$ 0.40 the two contributions become the same for $T =$ 70 K and below the value the coupling strength effect is larger than the pseudogap effect. $g$ is an important parameter for the discussion above.

We performed the same numerical analysis for our underdoped Y-123 with $T_c$ = 59 K. We display the results of analysis in figure \ref{fig3} and \ref{fig4}. The over all temperature dependent trends are very similar to those of Bi-2212 described previously. The peak position decreases almost linearly as temperature decreases as shown in the inset of the lower panel of figure\ref{fig3}. We compare the temperature dependent peak position in the electron-boson spectral function with that in the imaginary part of the local ($q$ averaged) magnetic susceptibility of underdoped Y-123 with $T_c$ = 52 K obtained by an inelastic neutron scattering\cite{fong00,fong00a}. We superimpose the two sets of the temperature dependent peak positions from both our optical and the neutron studies in the inset of the lower panel of figure \ref{fig3}. Interestingly, those two peaks agree almost perfectly in terms of energy scale and temperature dependence. This good agreement supports the proposal that the electron-boson density of cuprates has a magnetic origin, which would be the spin fluctuations\cite{carbotte99,hwang06,hwang07a,hwang08b}. In figure \ref{fig4} we can see that the pseudogap depth, $1 - \tilde{N}(0)$, is getting deeper as temperature decreases. The total spectral weight and the mass renormalization factor increase with similar temperature dependent trends as we observed in underdoped Bi-2212 system. $T_c^{MAX}(T)$ at all temperatures are higher than the actual $T_c$, 59 K; $T_c^{MAX}(T = 67 K)$ at our lowest temperature is 82 K, which is the lowest $T_c^{MAX}$ and about 40 \% higher than the actual $T_c$= 59 K. We also see a positive correlation between $T_c^{MAX}(T)$ and the peak position in $I^2\chi(\omega,T)$ but with a different proportionality, $\omega_{peak,s} \simeq 3.6 k_B T_c^{MAX}$. We note that this and previous relationships between the peak position and $T_c^{MAX}$ is similar to the well-known relationship between the magnetic resonance mode frequency and $T_c$, i.e., $\Omega_{res} \simeq 5.4 k_B T_c$ from inelastic neutron scattering \cite{he01}. The linear relationship between the resonance frequency and $T_c$ observed by other studies: $\Omega_{res} \simeq 4.9 k_B T_c$ from tunnelling\cite{zasadzinski01}, $\Omega_{res} \simeq 6 k_B T_c$ from ARPES\cite{johnson01}, and $\Omega_{res} \simeq 6.3 k_B T_c$ from optical studies\cite{yang09}.

\section{Conclusions}

The electron-boson spectral density function extracted from optical spectra of cuprates may carry crucial information about the bonding glue to form electron-electron Cooper pairs, which is the necessary process for superconductivity. It is not easy to determine uniquely the electron-boson density function from a given spectrum with a similar level of fitting quality\cite{dordevic05,schachinger06,heuman09}. However, those all different numerical methods give qualitatively similar results; the electron-boson density can be described by a couple of components. Those components are a temperature and doping dependent sharp mode, which is localized at low frequencies, and a relatively much less temperature and doping dependent broad mode, which spreads in a wide spectral range, up to 400 meV (3200 cm$^{-1}$) or higher. In our investigation using the new pseudogap model\cite{hwang08} we extracted the electron-boson spectral density function at various temperatures above $T_c$ from the optical self-energy of two underdoped cuprates: Bi-2212 and Y-123. The two different cuprates show common temperature dependent properties. The extracted electron-boson density function of underdoped cuprates shows a qualitatively different frequency dependence from those of optimally doped and overdoped cuprates\cite{hwang07a}. Most spectral weight of the resulting electron-boson density function is confined in low frequencies, within 1000 cm$^{-1}$. The sharp mode shifts to lower frequency as temperature decreases. This temperature dependent trend of the sharp mode is consistent with that of earlier inelastic neutron study\cite{fong00}.  We also observed that the pseudogap depth, $1-\tilde{N}(0,T)$, is getting deeper almost linearly and the spectral weight and the mass renormalization factor increases as temperature decreases. We estimated the fictitious (maximum) superconducting transition temperature, $T_c^{MAX}(T)$, from the extracted $I^2\chi(\omega,T)$ using the McMillan formula. It is always higher than the actual $T_c$ of the materials and interestingly decreases as temperature decreases. The reduction of $T_c^{MAX}$ with lowering temperature would be related to the pseudogap formation in underdoped cuprates, i.e., the pseudogap possibly suppresses the superconductivity in the materials.

%
%

\acknowledgments This work was supported for two years by Pusan National University Research Grant (2009-2011). I thank Dr. Timusk and Dr. Carbotte at McMaster University for their helpful discussions.

%
%


\begin{thebibliography}{14}
\bibitem{fong00a} We got the data points from figure 5a in Ref. \cite{fong00}; we took the peak position of the peak at the lower energy.
\end{thebibliography}
\end{document}